# Solid-Liquid Phase Transition As a Mechanism of Volcano Eruption

Alexander Ivanchin and Alexander Vikulin

This paper considers the formation of the magma volcano chamber and its eruption due to melting of the matter within the earth crust because of heating caused by plastic deformation occurring during tectonic movement. The expansion of matter in the magma chamber which takes place during its heating, leads to elastic stresses in the solid shell surrounding the magma chamber. The elastic energy of such stresses can be as high as 10^17 J per 1 km$^3$ of the melt. The magma flow rate has been assessed according to available data, which agrees well with the observation data. The mechanism of low-frequency vibrations produced by the magma chamber is discussed. The vibrations result from the excess elastic energy formed during melting at the eruption steady stage. The suggested radiation theory allows evaluating the size of the magma chamber according to parameters that can be measured. The obtained theoretical evaluation of the magma chamber size is supported by the available observation data.

## 1. Introduction

The existing theories describing a volcano eruption are based on degassing of magma (i.e. release of the gas dissolved in magma) arriving from the upper mantle of the Earth's depths (of the order of 100 km). All the theories contradict the seismology data [1]. In this study we consider the model of the formation and eruption of the magma volcano chamber due to melting of the matter in the earth crust (at depths up to 40 km) caused by the heat produced by plastic deformation during tectonic movement. That volcanoes have crust magma chambers is supported by the geological data [2].

From the physical point of view, the main issue concerning a volcano eruption is the problem of the formation of a region (an earth crust volume) with stored energy released during the eruption (explosion). The existing hypotheses of the formation of such regions are, as a rule, based on the idea that magma in the *liquid* state comes from the upper mantle and is stored in the *liquid* state in the magma chamber within the earth crust. The above approaches disagree with the available geophysical data [1], [3], [4]:

1. For a rather intensive (explosive) start of the volcano eruption, it is necessary that an area of increased pressure be formed in the earth crust below the volcano, i.e. a magma chamber. But how can such an area be formed inside the *liquid* magma?

2. The hypothesis of magma "floating up" from the mantle and entering the earth crust via the depth breaks is also confronted with a number of serious objections. Its weakest point is that magma does not cool down as it flows out onto the surface, although it takes quite a long time. We think that the various means that the authors use to prove their "floating up" hypothesis are not convincing from the physical point of view.

3. The volcano eruption is usually followed by magma flowing out onto the surface of the earth. Its flow rate gradually stabilizes and, in the course of time, decreases to zero. What is the reason for a decrease in the magma low rate? *Which pressure* fall in the magma chamber is related to the decrease in the magma flow rate?

4. Seismic waves do not reveal volumes containing the liquid phase below volcanoes. Lateral seismic waves attenuate, although they are not completely absorbed. The above data are a convincing evidence of the fact that below volcanoes there are no significant volumes containing the liquid phase, i.e. melted magma.

Also other problems arise if the reason for a volcano eruption is to be related to the liquid magma coming from the upper mantle from depths of the order of 100 km. Thus, the brief review shows that a volcano is not related to the mantle directly, and the volcano magma chamber capable of triggering volcano eruption are likely to form inside the earth crust. In this study we suggest the mechanism of the formation of a magma chamber located within the earth crust due to the solid-liquid phase transition as a result of plastic movement of the crust.

## 2. Heat and Melting

The characteristic periods of time between volcano eruptions are of the order of a thousand years. The depth of the earth crust is $L \sim 40$ km. Its characteristic thermal relaxation time is $t_* \sim L^2/a$, where $a$ is the thermal conductivity coefficient [5]. For rocks $a \sim 1\, mm^2/s$, then $t_* \sim$ 100,000 years. Therefore, one can neglect heat exchange in the first approximation. In this case, due to the plastic deformation $\varepsilon$ the temperature rises by $\Delta T = \tau \varepsilon / c\rho$. Here $\tau$ is the shear stress, c is the heat capacity, $\rho$ is the density.

Nearly for all solids the melting temperature rises with increasing pressure and the liquid phase density is less than that of the solid phase [6]. For simplicity, let us consider a heated region in the form of a sphere and put the origin of the spherical coordinates in its center.

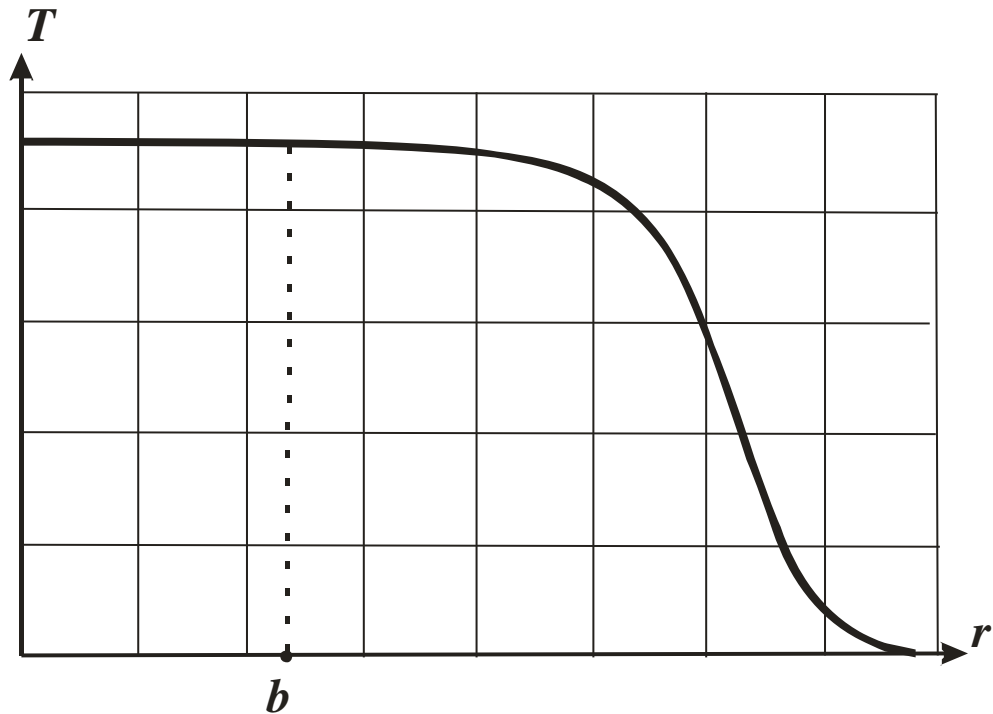

**Figure 1**

Figure 1 shows schematically the temperature profile in such a locally heated region. Point $b$ indicates the phase boundary. To the left of point $b$ there is the liquid phase and to the right – the solid phase. The hydrostatic pressure in the crust is

$$P = \rho g l \tag{2.1}$$

Here $\rho$ is the density, $g$ is the gravitational acceleration, $l$ is the depth. If the temperature in the heated region rises rather high, then the solid phase changes into the liquid phase with increasing volume. As the liquid phase is surrounded by the solid phase, there appears a surplus pressure $\Delta p$ that produces an elastic field in the solid phase [7]

$$\mathbf{u} = \frac{C}{r^2}\{1,0,0\} \tag{2.2}$$

Here $C$ is an arbitrary constant, and the order of the coordinates in the spherical system is the following: radial, zenith and azimuthal. The deformation and stress tensors in the spherical coordinates are written as

$$\varepsilon_{rr} = -\frac{2C}{r^3}, \quad \varepsilon_{\theta\theta} = \varepsilon_{\varphi\varphi} = \frac{C}{r^3}$$

$$\sigma_{rr} = -\frac{2CG}{r^3}, \quad \sigma_{\theta\theta} = \sigma_{\varphi\varphi} = \frac{2CG}{r^3} \tag{2.3}$$

The other tensor components are equal to zero. Here $G$ is the shear modulus. At the phase boundary at $r=b$ the stress $\sigma_{rr}$ normal to the boundary is equal to the liquid pressure

$$\sigma_{rr} = -\frac{2CG}{r^3} = \Delta p \tag{2.4}$$

At constant temperature in two-phase equilibrium systems the pressure $\Delta p$ does not depend on the phase volume. It holds true under the condition

$$C = \alpha b^3 \tag{2.5}$$

Here $\alpha$ is the proportionality coefficient. In this case, (2.4) will change into $\Delta p = -2G\alpha$. According to (2.2) and (2.5), the elastic displacement at the phase boundary is $u_r = \alpha b$. Due to that an increase in the liquid volume will be $4\pi\alpha b^3$. Dividing the above value by the liquid sphere volume $4\pi b^3/3$ we obtain the relative change of the liquid sphere volume $3\alpha$. Therefore, the liquid phase pressure will be

$$\Delta p = \kappa(\omega - 3\alpha) \tag{2.6}$$

Here $\kappa$ is the volume liquid elastic modulus, $\omega = (\rho - \rho_l)/\rho$ is the relative change of volume at melting, $\rho_l$ is the liquid phase density, $\rho$ is the solid phase density. The total pressure taking into account (2.1) is written as:

$$P = \rho g l + \Delta p \tag{2.7}$$

From (2.4) and (2.6) we obtain

$$\alpha = \frac{\kappa\omega}{3\kappa - 2G}$$

For real materials the relation $\kappa/G \sim 1$, therefore $\alpha \sim \omega$ and further suppose $\alpha = \omega$. The stresses (2.3) are written as

$$\sigma_{rr} = -\frac{2G\omega b^3}{r^3}, \quad \sigma_{\theta\theta} = \sigma_{\varphi\varphi} = \frac{G\omega b^3}{r^3} \tag{2.8}$$

and the pressure (2.6) as

$$\Delta p = -2\kappa\omega \tag{2.9}$$

Here the sign of minus corresponds to the compression stresses

The elastic field energy of the solid phase is [7]:

$$W_s = G \iiint_V (\varepsilon_{rr}^2 + \varepsilon_{\theta\theta}^2 + \varepsilon_{\varphi\varphi}^2) dV = 6\pi b^3 G \omega^2 \qquad (2.10)$$

Integration is carried out over the solid phase that is for $r > b$. The elasticity energy of the liquid region over (2.9) is

$$W_l = 2\pi b^3 \frac{(\Delta p)^2}{\kappa} = 8\pi b^3 \kappa \omega^2 \qquad (2.11)$$

Thus, the elastic energies of the liquid region and the solid shell caused by the phase transition are values of the same order. The total elastic energy is

$$W = W_s + W_l = 14\pi b^3 \kappa \omega^2$$

Equating (2.10) to (2.11) we obtain a condition of equality of the elastic energies in the form of $= 3G/4$. If $< 3G/4$, then the elastic energy of the liquid phase is less than that of the solid phase. If $\kappa > 3G/4$, then it is vice versa.

The process develops in the following way. As a result of the heat release due to the plastic deformation, a locally heated region is formed in the earth crust. When the temperature of melting is reached, there appears the liquid phase. The heat needed for melting comes from the solid phase. As the solid phase is heated up, the liquid phase $b$ grows in size as well as the stresses (2.8) and the energy (2.10) until an explosion occurs. At the explosion the surplus pressure in the volcano chamber drops abruptly down to the value $\Delta p_1 < \Delta p$, which leads to a jump-like transition of the solid phase surrounding the liquid volcano chamber into the liquid state with growing specific volume, which increases considerably the explosive force. The above scenario is absolutely analogous with what happened at the explosion of the boilers in the 19[th] century, in which overheated water was in the liquid state under high pressure. As a result of the boiler depressurization, the water jump-like turned into steam leading to an explosion of a great force. After the explosion the total pressure in the volcano chamber is

$$P = \rho g l + \Delta p_1 \qquad (2.12)$$

and then a steady stage of the volcano eruption begins, i.e. a magma outflow. This stage can start even without a previous explosion if there are no serious obstacles for a magma outflow. When a substance changes from the solid state to the liquid state in a jump-like way, its temperature in the liquid state is lower than that in the solid state near the phase boundary, because during melting heat is absorbed. Due to thermal conductivity the heat from the solid phase passes to the liquid phase. Besides, there is also another process under way. Since magma leaves the volcano chamber, the pressure $\Delta p_1$ must decrease. However, it does not, because the change of the solid phase to the liquid phase is accompanied by increasing specific volume. As is known from the laws of the phase equilibrium of two-phase one-component systems [6], removal of one of the phases leads to the phase transition of part of the matter, with the system temperature and pressure remaining the same. As a result, the phase boundary moves inside the solid phase increasing the volume of the liquid phase. The above process sustains a steady magma outflow. However, as the temperature decreases with the advance of the phase boundary, the process gradually slows down to a complete stop. As a result, the liquid will be heated nonuniformly, with its temperature near the volcano chamber center being higher and equal to that of melting at the pressure $\Delta p + \rho g l$. The solid phase temperature near the phase boundary will also be higher than that of the liquid adjacent to the boundary. It takes the time of the order of $b^2/a$ for the temperature to become uniform due to diffusion thermal conductivity. For $b \sim 1$ km the time is $\sim 1000$ years. Near the solid phase the temperature is close to that of melting at the pressure $\Delta p_1 + \rho g l$ and it will be lower than that of the surrounding solid phase. The heat from the solid phase passes to the liquid phase, which results in a complicated temperature profile, because the thermal relaxation time is very long.

According to mechanics, the flow of viscous liquid in a cylindrical tube is characterized by the relation [8]

$$Q = \frac{\pi \chi^4 \Delta p_1}{8 \mu l}$$

Here $Q$ is the volume magma flow rate per second through the volcanic pipe, $\chi$ is the volcano channel (uptake) radius, $\mu$ is the magma viscosity, $l$ is the depth at which the volcano chamber occurs (the volcano channel length from the surface to the volcano chamber). Due to melting the phase boundary will move a distance of $dr = \eta dt$ during the time period $dt$. Here $\eta$ is the rate of movement of the phase boundary. The liquid sphere volume will increase by $dV = 4\pi b^2 \eta dt$. The part of the solid matter $\rho dV$ will turn into the liquid state. The excess volume of the liquid equal to $\omega dV$ formed at melting due to expansion has to be removed from the volcano chamber to avoid increasing the pressure. We obtain the volume magma flow rate in the form

$$\dot{V} = \frac{dV}{dt} = 4\pi \omega b^2 \eta \qquad (2.13)$$

For $\rho \sim 3000 \; kg/m^3$, $\omega \sim 0{,}05$, $b \sim 1$ км, $\eta \sim 1 \; mm/s$ we obtain $\dot{V} \sim 600$ м$^3$/с.

We evaluate the energy released during an eruption using aluminum as an example, since we could not find the necessary thermodynamic basalt parameters. Its melting temperature is $930K$, and its thermodynamic characteristics have been examined thoroughly due to technical importance. The temperature of liquid lava begins from $1000K$ and is close to the aluminum melting temperature. According to [9], the aluminum melting temperature versus the pressure is $T_* = 9 \cdot 10^{-8} p + 940$. Here the pressure is expressed in Pascal's and the temperature in Kelvin degrees. The phase transition heat is $10.8 \; kJ/mol$. An increase in the specific volume at melting is $\Delta V = 0.724 \; cm^3/mol$, which corresponds to the relative expansion $\omega = 0{,}064$. The shear modulus for aluminum is $G = 2.45 \cdot 10^{10} \; Pa$ [10]. At a pressure of 60 kbar the relative compression value for aluminum is 0.08. The elastic field energy (2.10) for $b = 1$ km is of the order of $10^{17} J$, that is, 50 megatons in the trotyl equivalent. The heat capacity of aluminum at $1000K$ is $30 \; J/mol \cdot K$. Hence it follows that under atmospheric pressure at heating from a melting temperature of $980 \; K$ up to a temperature of $1300 \; K$, the amount of heat consumed will be same as during the phase transition. At a temperature of $1300 \; K$ and a pressure of $6 \cdot 10^9 \; Pa$ aluminum is in the solid phase. If the pressure falls, it will turn into the liquid phase abruptly with the corresponding release of energy.

### 3. Low-Frequency Volcano Radiation

After the volcano eruption low-frequency vibrations occur which are an infra-sound of a frequency of the order of $1 - 10 \; s^{-1}$ [11]. Possibly its radiation mechanism at the steady stage is as follows. Let the liquid region as a whole undergo radial vibrations of the frequency $\nu$, with the period $1/\nu$ greatly exceeding the time within which the sound passes the liquid sphere $2b/c$, here $c$ is the sound speed in the liquid. In this case, the liquid pressure can be considered the same throughout its volume and depending only on time. We add to the pressure (2.12) the periodic sound pressure $\Pi \cos(2\pi\nu t)$, where $t$ is the time, $\Pi \ll \Delta p_1$ is the amplitude. The total pressure is

$$P = \Delta p_1 + \rho g l + \Pi \cos(2\pi\nu t)$$

At the moments of time $P = \Delta p_1 + \rho g l + \Pi \cos(2\pi\nu t)$ the cosine takes the value of $-1$, the sound pressure is equal to $-\Pi$, the liquid volcano chamber undergoes extension, the relative decrease in the liquid pressure is $\Pi/\kappa$ the displacement of the phase boundary from the equilibrium position is $3b\Pi/\kappa$ also a decrease in the density of the liquid potential energy caused by the extension $\Pi^2/2\kappa$. Multiplying that value by the liquid volume we get the maximum value of decreasing potential energy of the liquid volcano chamber at the extension $2\pi b^3 \Pi^2/2\kappa$. In some time equal to one period the volcano chamber reaches the state of the maximum extension again. However, due to the volcano chamber extension caused by melting, the size of the liquid volcano chamber within one period $1/\nu$ will increase by $db = \eta/\nu$. The elastic energy of the solid phase from the sphere layer will change into the liquid elastic energy. If the condition $\kappa < 3G/4$ is satisfied, then the elastic energy of the liquid formed will be larger than that of the melted solid by

$$6\pi b^2 \omega^2 (3G - 4\kappa)\frac{\eta}{\nu} \tag{3.1}$$

and the surplus will be radiated in the form of the elastic wave energy into the complete solid angle. Multiplying (3.1) by the frequency we derive the radiation power in the form of $6\pi b^2 \omega^2 (3G - 4\kappa)\eta$. In this way the sounding is supplied with energy. The volcanic sounding can be recorded during rather a long period of time - for several days and even longer [11].

The wave equation describing the sounding is

$$\frac{\partial^2 u}{\partial t^2} = c_s^2 \Delta u \tag{3.2}$$

Here $\Delta$ is the Laplace operator and $c_s$ is the sound speed in the solid phase. As the problem is centrally symmetric, there is only one displacement radical component left which is denoted by $u$, and it depends only on the distance to the origin of the coordinates. The absence of the zenith component $u_\theta$ and the azimuth component $u_\varphi$ in the displacement in the spherical coordinate system results from the fact that the liquid cannot produce shifts at the phase boundary tangential to the boundary. The only component that can be produced due to pressure is the displacement $u$ normal to the boundary. Hence it follows that the volcanic sound can have only a longitudinal wave. There must be no shift component in the volcanic radiation.

The boundary condition at the phase boundary at $r = b$ is $p = \Pi \cos(2\pi\nu t)$. The solution (3.2) is [12]

$$u = \frac{\Pi b}{r} \cos 2\pi \left(\nu t - \frac{r}{\lambda}\right) \tag{3.3}$$

Here $\lambda = c_s/\nu$ is the wavelength. The elastic deformation is

$$\varepsilon_{rr} = \frac{\partial u}{\partial r} = 2\pi b \Pi \left[ \frac{1}{\lambda r} \sin 2\pi \left( vt - \frac{r}{\lambda} \right) - \frac{1}{r^2} \cos 2\pi \left( vt - \frac{r}{\lambda} \right) \right]$$

(3.4)

$$\varepsilon_{\theta\theta} = \varepsilon_{\varphi\varphi} = \frac{\Pi b}{r^2} \cos 2\pi \left( vt - \frac{r}{\lambda} \right)$$

The other components of the deformation tensor are equal to zero. It follows from (3.4) that at distances $r \gg \lambda$ (in the far zone) in $\varepsilon_{rr}$ only the first term in the square brackets is left, while $\varepsilon_{\theta\theta}$ and $\varepsilon_{\varphi\varphi}$ disappear. In this case, the deformation has only one non-zero component left

$$\varepsilon_{rr} = \frac{\partial u}{\partial r} = 2\pi \frac{\Pi b}{\lambda r} \sin 2\pi \left( vt - \frac{r}{\lambda} \right) \tag{3.5}$$

The elastic displacement rate is written as

$$\dot{u} = \frac{\partial u}{\partial t} = -2\pi \Pi v \frac{b}{r} \sin 2\pi \left( vt - \frac{r}{\lambda} \right)$$

Then the kinetic energy in the spherical layer of the wave length depth in the complete solid angle looks like

$$W_K = 2\pi\rho \int_r^{r+\lambda} \dot{u}^2 r^2 dr = 4\pi^3 \rho (Abv)^2 \lambda = 4\pi^3 \rho (\Pi b)^2 cv$$

The kinetic energy must be equal to that radiated by the liquid nucleus (3.1). As a result, we derive

$$\eta = \frac{2\pi^2 \rho \Pi^2 c v^2}{\omega^2 (3G - 4\kappa)} \tag{3.6}$$

Excluding the parameter $\eta$ from (2.13) and (3.6) we get

$$b = \frac{1}{2\pi \Pi v} \sqrt{\frac{\dot{V}(3G - 4\kappa)}{2\pi \rho c}} \tag{3.7}$$

The formula expresses the dimension of the volcano chamber through the values that can, in principle, be measured at the modern scientific level. For example, let us carry out evaluation using the following values of the parameters $\rho = 3 \ ton/m^3$, $c = 5 \ km/s$, $v = 10 \ s^{-1}$, $\omega = 0.05$, $\Pi = 1 \ mm$, $\dot{V} = 100 \ m^3/s$, $3G - 4\kappa = 10^9$ Па. According to (3.7) we obtain $b = 900 \ m$. The above value is close to the size of the magma chamber found during the eruption of volcano Tolbachik in 1975-1976 [13].

## Conclusion

One of the problems in volcanology is how localization of surplus energy needed for explosion occurs. One can suppose that its mechanism is unlikely to form in the mantle magma (liquid). Even if it could form there, one has to imagine the mechanism of melting of the lithosphere from the magma to the surface so that the volume with the energy concentration would rise to the Earth's surface. The mechanism of heating up taking into account melting solves the above problem. Local heating is caused by plastic deformation occurring during tectonic movement and brings about a change of the solid phase into the liquid. The numerical evaluation performed shows that the suggested theory agrees with the observation data on real volcanoes. The mechanism of producing sound vibrations caused by viscosity during the magma flow was considered in [14]. However, that mechanism is implemented if there is a peculiar dependence of viscosity on the flow rate. We do not know real liquids possessing such viscous properties.